\documentclass[a4paper,preprintnumbers,amsmath,amssymb,twocolumn,10pt]{revtex4-2}

\usepackage{graphicx}% Include figure files

\usepackage{dcolumn}% Align table columns on decimal point
\usepackage{bm}% bold math
\usepackage{epstopdf}
\newcount\driver
\newcount\bozza

scaled\magstep1                  
scaled\magstep1 
 
scaled\magstep1                 \font\indbf=cmbx10 scaled\magstep2

scaled \magstep2

{\count255=\time\divide\count255 by 60
\xdef\hourmin{\number\count255}
        \multiply\count255 by-60\advance\count255 by\time
   \xdef\hourmin{\hourmin:\ifnum\count255<10 0\fi\the\count255}}

\let\a=\alpha \let\b=\beta    \let\g=\gamma     \let\d=\delta     \let\e=\varepsilon
       \let\th=\vartheta      \let\l=\lambda
\let\m=\mu    \let\n=\nu                      
\let\s=\sigma \let\t=\tau            
\let\ps=\psi        
\let\G=\Gamma        \let\L=\Lambda

\def\EE{{\cal E}}\def\VV{{\cal V}}

\def\TT{{\cal T}}\def\ZZ{{\cal Z}}
\def\RR{{\cal R}}\def\LL{{\cal L}}

\def\nn{{\bf n}}

       \def\oo{{\underline \omega}}
\def\ee{{\underline \varepsilon}}

        \def\EE{\hbox{\msytw E}}

\let\==\equiv

\let\io=\infty
\let\0=\noindent

\def\*{{\hfill\break\null\hfill\break}}

\def\tilde#1{{\widetilde #1}}

\def\tende#1{\,\vtop{\ialign{##\crcr\rightarrowfill\crcr
             \noalign{\kern-1pt\nointerlineskip}
             \hskip3.pt${\scriptstyle #1}$\hskip3.pt\crcr}}\,}
\def\otto{\,{\kern-1.truept\leftarrow\kern-5.truept\to\kern-1.truept}\,}

\def\wh#1{\widehat{#1}}
\def\hat#1{\wh{#1}}
\def\sqt[#1]#2{\root #1\of {#2}}

\def\bp{{\bar \ps}}

\def\EE{{\cal E}}\def\VV{{\cal V}}

\def\TT{{\cal T}}\def\ZZ{{\cal Z}}
\def\RR{{\cal R}}\def\LL{{\cal L}}

\def\T#1{{#1_{\kern-3pt\lower7pt\hbox{$\widetilde{}$}}\kern3pt}}
\def\VVV#1{{\underline #1}_{\kern-3pt
\lower7pt\hbox{$\widetilde{}$}}\kern3pt\,}
\def\W#1{#1_{\kern-3pt\lower7.5pt\hbox{$\widetilde{}$}}\kern2pt\,}

\def\indica{\leaders \hbox to 0.5cm{\hss.\hss}\hfill}
\def\guida{\leaders\hbox to 1em{\hss.\hss}\hfill}
\mathchardef\oo= "0521

\def\nn{{\bf n}}

\def\oo{{\underline \omega}}

\def\qed{\raise1pt\hbox{\vrule height5pt width5pt depth0pt}}
 
  \def\bp{{\bar p}} 

\def\indic{\hbox{\raise-2pt \hbox{\indbf 1}}}

\def\ins#1#2#3{\vbox to0pt{\kern-#2 \hbox{\kern#1 #3}\vss}\nointerlineskip}

\newdimen\xshift \newdimen\xwidth \newdimen\yshift
\def\insertplot#1#2#3#4#5#6{%
\xwidth=#1pt \xshift=\hsize \advance\xshift by-\xwidth \divide\xshift by 2%
\begin{figure}[ht]
\vspace{#2pt} \hspace{\xshift}
\begin{minipage}{#1pt}
#3 \ifnum\driver=1 \griglia=#6
\ifnum\griglia=1 \openout13=griglia.ps \write13{gsave .2
setlinewidth} \write13{0 10 #1 {dup 0 moveto #2 lineto } for}
\write13{0 10 #2 {dup 0 exch moveto #1 exch lineto } for}
\write13{stroke} \write13{.5 setlinewidth} \write13{0 50 #1 {dup 0
moveto #2 lineto } for} \write13{0 50 #2 {dup 0 exch moveto #1
exch lineto } for} \write13{stroke grestore} \closeout13
\includegraphics{griglia.ps} \fi
\includegraphics{#4.ps}\fi%
\ifnum\driver=2 \fi
\end{minipage}
\caption{#5}
\end{figure}
}
\newdimen\shift \shift=-1.5truecm
\def\lb#1{%
\ifnum\bozza=1
\label{#1}\rlap{\hbox{\hskip\shift$\scriptstyle#1$}}
\else\label{#1} \fi}

\def\be{\begin{equation}}
\def\ee{\end{equation}}
\def\bea{\begin{eqnarray}}\def\eea{\end{eqnarray}}
\def\bean{\begin{eqnarray*}}\def\eean{\end{eqnarray*}}
\def\bfr{\begin{flushright}}\def\efr{\end{flushright}}
\def\bc{\begin{center}}\def\ec{\end{center}}
\def\bal{\begin{align}}\def\eal{\end{align}}
\def\ba#1{\begin{array}{#1}} \def\ea{\end{array}}
\def\bd{\begin{description}}\def\ed{\end{description}}
\def\bv{\begin{verbatim}}\def\ev{\end{verbatim}}
\def\nn{\nonumber}
\def\Halmos{\hfill\vrule height10pt width4pt depth2pt \par\hbox to \hsize{}}
\def\pref#1{(\ref{#1})}

\def\ins#1#2#3{\vbox to0pt{\kern-#2 \hbox{\kern#1 #3}\vss}\nointerlineskip}

\newdimen\xshift \newdimen\xwidth \newdimen\yshift
\newcount\griglia

\def\insertplot#1#2#3#4#5#6{%
\xwidth=#1pt \xshift=\hsize \advance\xshift by-\xwidth \divide\xshift by 2%
\begin{figure}[ht]
\vspace{#2pt} \hspace{\xshift}
\begin{minipage}{#1pt}
#3 \ifnum\driver=1 \griglia=#6
\ifnum\griglia=1 \openout13=griglia.ps \write13{gsave .2
setlinewidth} \write13{0 10 #1 {dup 0 moveto #2 lineto } for}
\write13{0 10 #2 {dup 0 exch moveto #1 exch lineto } for}
\write13{stroke} \write13{.5 setlinewidth} \write13{0 50 #1 {dup 0
moveto #2 lineto } for} \write13{0 50 #2 {dup 0 exch moveto #1
exch lineto } for} \write13{stroke grestore} \closeout13
\includegraphics{griglia.ps} \fi
\includegraphics{#4.ps}\fi%
\ifnum\driver=2 \fi
\end{minipage}
\caption{#5}
\end{figure}
}
\newdimen\shift \shift=-1.5truecm
\def\lb#1{%
\label{#1}\rlap{\hbox{\hskip\shift$\scriptstyle#1$}}
\else\label{#1} \fi}

\def\be{\begin{equation}}
\def\ee{\end{equation}}
\def\bea{\begin{eqnarray}}\def\eea{\end{eqnarray}}
\def\bean{\begin{eqnarray*}}\def\eean{\end{eqnarray*}}
\def\bfr{\begin{flushright}}\def\efr{\end{flushright}}
\def\bc{\begin{center}}\def\ec{\end{center}}
\def\bal{\begin{align}}\def\eal{\end{align}}
\def\ba#1{\begin{array}{#1}} \def\ea{\end{array}}
\def\bd{\begin{description}}\def\ed{\end{description}}
\def\bv{\begin{verbatim}}\def\ev{\end{verbatim}}
\def\nn{\nonumber}
\def\Halmos{\hfill\vrule height10pt width4pt depth2pt \par\hbox to \hsize{}}
\def\pref#1{(\ref{#1})}

\usepackage{amsmath}
\usepackage{amsfonts}
\usepackage{amssymb}
\usepackage{epstopdf}

scaled\magstep1

\let\a=\alpha \let\b=\beta  \let\g=\gamma  \let\d=\delta
\let\e=\varepsilon
     \let\th=\theta  \let\l=\lambda
\let\m=\mu    \let\n=\nu             
\let\s=\sigma \let\t=\tau    
\let\ps=\Psi   
\let\G=\Gamma   \let\L=\Lambda

\def\EE{{\cal E}} \def\VV{{\cal V}}
 
\def\TT{{\cal T}} 
\def\RR{{\cal R}}\def\LL{{\cal L}}

\def\nn{\nonumber}

\def\\{\hfill\break}
\def\={:=}
\let\io=\infty
\let\0=\noindent

\def\tende#1{\,\vtop{\ialign{##\crcr\rightarrowfill\crcr\noalign{\kern-1pt
    \nointerlineskip} \hskip3.pt${\scriptstyle #1}$\hskip3.pt\crcr}}\,}
\def\otto{\,{\kern-1.truept\leftarrow\kern-5.truept\to\kern-1.truept}\,}

\def\wh{\widehat}
\def\to{\rightarrow}

\def\qed{\hfill\raise1pt\hbox{\vrule height5pt width5pt depth0pt}}

\def\be{\begin{equation}}
\def\ee{\end{equation}}
\def\bp{\begin{pmatrix}}
\def\ep{\end{pmatrix}}
\def\bea{\begin{eqnarray}}
\def\eea{\end{eqnarray}}
\def\nn{\nonumber}
\def\pref#1{(\ref{#1})}

\def\lb{\label}

\begin{document}

\title{Non-perturbative RG for the Weak interaction corrections to
the magnetic moment}

\author{Vieri Mastropietro}

\affiliation{University of Milan, Italy}

\begin{abstract} 
We analyze, by rigorous Renormalization Group (RG) methods,
a Fermi model for Weak forces with a 
single family of leptons,
one massless and the other with mass $m=M e^{-\b}$, with $M$ the gauge boson mass,
a quartic non-local interaction with coupling $\l^2$ and a momentum cut-off $\L$.
The 
magnetic moment is written as a series in $\l^2$, with 
$n$-th coefficients 
bounded by $C^n  
({m^2\over M^2}) \b^{2n }
({\L^2\over M^2})^{(1+0^+)(n-1)}$ if $C$ a constant; this 
implies convergence and provides non-perturbative bounds on the higher orders contribution.
%The condition  $M/\L$ small respect to $1$ ensures that 
%lower orders are insensitive to the cut-off.
The fact that the magnetic moment is associated to a dimensionally irrelevant quantity requires the implementation of cancellations
in the multiscale analysis.
\end{abstract} 

\maketitle

%\section{Effective electroweak theory and main result}

\section{Introduction and main results}  

The anomalous magnetic moment ${g-2\over 2}$ plays a central role in physics since the beginning of Quantum Field Theory  \cite{2}
and it is nowadays attracting a renewed interest \cite{mu}.
Its theoretical value 
can be computed in the Standard Model with very high precision and
comparison with experiments provides a stringent test on the completeness of the theory.

The contributions to the anomalous magnetic moment can be divided in the ones involving 
also Strong forces and the ones considering only
ElectroWeak ones.
In the first case the non-perturbative nature of the  low energy Strong interactions requires numerical lattice or data driven approaches, see 
e.g \cite{1aa}, \cite{11aa}, \cite{3aa}. 
In the second, an analytical perturbative approach is in principle justified 
by the smallness (in adimensional units)
of the couplings involved,
that is $\a= 1/137,..$ and $\l^2=4\pi\a/\sin\th_W^2$ ($\sin^2 \th_W=0,2231..)$.
%1-({M_W\over M_Z})^2=0,231..$);
%The uncertainty comes in this case from the truncation of the series expansion, that is from 
%the size of the neglected infinitely many higher order terms. 

The ElectroWeak theory allows to write the magnetic moment as a series
$\sum
\a^n \l^{2m}   A_{n,m}$ with coefficients $A_{n,m}$ expressed by sum of Feynman graphs. 
Perturbative renormalizability \cite{tH} (see also \cite{KM}), ensures that the ultraviolet divergences present in the graphs can be exactly compensated by a suitable choice of the bare parameters, so that each coefficient $A_{n,m}$ is finite removing the cut-offs, typically with a factorial growth in the order. 

The
coefficients $A_{n,m}$ can be explicitly computed and their evaluation becomes more and more challenging increasing the order.
In the case of the pure QED contributions $A_{n,0}$ the first order was computed in \cite{2} $A_{1,0}=1/2\pi$,
the second in \cite{3} and more recent computations were done up to $n=5$, see eg  \cite{4},  \cite{5} and the review \cite{3aa}.
Such coefficients are universal numbers (if a single lepton is considered) 
depending only on $\a$.
In contrast, the Weak-interaction corrections depend on the lepton masses; in particular, see  \cite{6},  $A_{0,1}={5\over 24\sqrt{2}\pi^2}{m^2\over M_W^2}  (1+{1\over 5}(1-4s^2)^2)$
%with $s^2=1-({M_W\over M_Z})^2$,
$m$ the lepton mass and $M_W$ the $W$ mass. The smallness of the ratio
$(m/M_W)^2$
says that the Weak contributions are quite suppressed with respect to the e.m. ones, in particular in the case of the electron.  

While such perturbative computations were in spectacular agreement with measurements, the 
increasing level of experimental accuracy and some apparent discrepancy in the muon case
provides a motivation to reconsider the role of higher order terms.
Indeed the predictions are done
{\it truncating} the series expansion at a certain order $n$, and the effect of higher orders is estimated to be 
$\a^{n+1}$ in the case of QED or ${m^2\over M_W^2}\l^{2(n+1)}$
for Weak forces, up to a constant $C^n$ with $C$ of size
suggested by lowest orders.
However such series are not convergent (if so the error would be indeed $O(C^n \e^n)$
if $\e$ is the coupling), so that the truncation cannot be done at arbitrary order; if asymptotic
the error would be $O(C^n n! \e^n)$ (and the truncation could be done only up to a finite order), but it is likely 
that, at least if one restricts to the ElectroWeak sector,
even this is not the case \cite{a}, \cite{a1}
due to the triviality phenomenon, 
rigorously established for $\phi^4$  \cite{b}, \cite{c}. Other sources of non-perturbative errors in the truncation are in 
\cite{12}, \cite{13}. 

The magnetic moment is used as an experimental input to get the value of the fine structure constant, as
is done also for other quantities like the Hall conductivity, \cite{13a}, \cite{13b}. However for the latter there is no theoretical uncertainity due to truncation: even if in principle it could acquire corrections 
due the presence of many body interactions, all higher orders are {\it
exactly vanishing} due to topological protections, as recently rigorously established \cite{13c}, \cite{13dd} .
This is however {\it not } the case for the anomalous magnetic moment, and an estimate on the higher order terms neglected in the perturbative approach is therefore necessary.

A {\it non-perturbative } framework is obtained by expressing the magnetic moment in terms of functional integrals regularized
with a finite ultraviolet cut-off $\L$ in the Euclidean setting, which is suitable for the magnetic moment \cite{13}.
The cut-off
must be much larger than the experiments scale, so that the
results are expected to be cut-off independent. 
On the other hand, the cut-off cannot be taken arbitrarily high, at least if one considers only ElectroWeak forces, due to the triviality.
There is a well known relation between the renormalizability properties and the maximal allowed cut-off.
In a renormalizable model, like the ElectroWeak sector,
one expects in principle that a cut-off at least exponentially high in the inverse coupling can be reached ensuring, due to the smallness of the coupling,
that cut-off corrections are negligible. This however requires 
a non-perturbative formulation 
of the ElectroWeak theory and there are well known difficulties for a chiral Gauge theory like that  \cite{lat},\cite{lat1},\cite{lat2}.

We consider therefore lower values of the cut-off $\L$ where 
the Weak forces can be described by Fermi interactions, and we restrict to
a single family of leptons $l,\n$. 
%The analysis is based on non-perturbative RG.
%Such methods were previously applied to the chiral anomaly \cite{M11}, \cite{M11a}, \cite{M12}, but the anomalous magnetic moment is associated to 
%{\it irrelevant} terms the RG sense (not marginal like the anomaly), and this requires the implementation of suitable cancellations in the multiscale analysis.
The
{\it Effective potential} is given by
\be
e^{V^{e}_{\L}(A,\phi)}=\int P(d\psi)e^{V(\psi+\phi)+\bar B(A,\psi+\phi)}\label{ippo}
\ee 
where $\psi_{x,i},\bar\psi_{x,i}$ are Grassmann variables, $i=l, \n$ is the particle
index,
$x\in (0,L]^4$ and periodic boundary conditions are imposed, $\psi_{x,i}=(\psi^-_{x,i,L},\psi^-_{x,i,R})$, $\bar\psi_{x,i}=(\psi^+_{x,i,R},
\psi^+_{x,i,L})$,
$\g_0= \begin{pmatrix} 0 & I \\ I &0 \end{pmatrix}\quad \g_j= \begin{pmatrix} 0 & i\s_j \\-i\s_j &0 \end{pmatrix}$,
% \quad\g_5=\begin{pmatrix}&I&0\\
%          &0&-I\end{pmatrix}$. 
and $P(d\psi)$ is the fermionic integration
with propagator, $i=l,\n$ \be g_i(x-y)={1\over L^4}\sum_k e^{i k(x-y)}
{\chi_N(k)\over \not k+m_i}\ee where
%with $\tilde \g_0= \begin{pmatrix} 0 & I Z_{N,L,i} \\ I  Z_{N,R,i}&0 \end{pmatrix}\quad \g_j= \begin{pmatrix} 0 & i\s_j Z_{N,L,i}\\-i\s_j Z_{N,R,i}&0 \end{pmatrix}%$ and  $\tilde m_i= \begin{pmatrix} \sqrt{Z_{N,L,i}Z_{N,L,i} } m_i & 0 \\ 0}&\sqrt{Z_{N,L,i}Z_{N,L,i} } m_i \end{pmatrix}$.
%The parameters $Z_{N,s,i}$ are chosen so that the dressed wave function renormalization is $=1$ and the dressed mass is $m_i$.
 $\chi_N(k)=\chi(\g^{-N} k)$, with $\chi(k)$ is a cut-off function
 such that $\chi=1$ for $|k|\le 1/\g$
and $\chi=0$
for $|k|\ge 1$ and $\L=\g^N$, with $N$ a positive integer, $\g>1$ is a scaling parameter. Moreover $\s_\m^L=(\s_0,i \s) $ and $\s_\m^R=(\s_0,-i \s) $ with $\s_1=\begin{pmatrix}&0&1\\
          &1&0\end{pmatrix}
 \quad
\s_2=\begin{pmatrix}
&0&-i\\ &i&0
\end{pmatrix}
\quad\s_3=\begin{pmatrix}&1&0\\
          &0&-1\end{pmatrix}$. 
The interaction is given by 
\be
V=
{\l^2 \over 2 }  \int dx  d y [v_W(x,y) j_{\m,x}^{+W}  j_{\m,y}^{-W} + v_Z(x,y) j_{\m,x}^Z  j_{\m, y}^Z]\label{quar} 
\ee
%
%With a suitable $\l$ it becomes $G$; we want to fix $\g^N=M_Z$, 
%that is the $Z$ scale; $\l$ is the \quare of electric charge.
%with $v_\a(x,y)=\int dk e^{i k(x-y)}
%${\chi(k)\over k^2+M_\a^2 }$, $\a=W,Z$
with $\hat v_W(k)={\chi_N(k)\over k^2+M_W^2}$, $\hat v_Z(k)={\chi_N(k)\over k^2+M_Z^2}$,
the charged currents are $j_{\m,x}^{+ W}= 
\psi^+_{l,L,x} \s^L_\m \psi^-_{\nu,L,x}$,
$j_{\m,x}^{+ W}=\psi^+_{\n,L,x}\s^L_\m \psi^-_{l,L,x}$
and the neutral current is, $s=L,R$ 
\be
j^Z_{\m,x}=\sum_{i,s}(\e_s-\sin^2 \th_W Q_i)
\psi^+_{x,i,s}\s_\m^s \psi^-_{x,i,s}
\ee 
with $s=L,R$
$Q_l=Q, Q_\n=0$ and 
$\e_L=-\e_R=1$. Note that the interaction is non local in space and it decays with inverse rate $M_W,M_Z$. 

The source term is given by
$\bar B(A,\psi)=\int dx A_\m j^{e.m.}_{\m,x}$
%+
%\int dx 
%(\bar\psi_x\phi_x+\psi_x\bar\phi_x)$ 
with  $j^{e.m.}_{\m,x}$ the total e.m. current
$
j^{e.m.}_x=\sum_{s} \ZZ_{s}   Q\psi^+_{x,l,s}  \s^s_\m \psi^-_{x,l,s}$.
The fermion $l$ is massive and the fermion $\n$ massless, $m_l=m$ and $m_\n=0$; moreover we define $M_W=M, M_Z=\cos \th_W M$ with
$\cos\th_W\sim 0,881...$.
and $m/M=e^{-\b}$ with $\b$ $\sim 3$ for muons and $\sim 6$ for electrons, and $\L\ge M$. 

The {\it vertex function} is given by
$\G_{\m,i,s,s'}(z;x,y) ={\partial^3 V^e_{\L}\over \partial A_{\m,z} \partial \phi^-_{x,i,s}\phi^+_{y,i,s'}}|_0$. If $\hat\G_{\m,i,s,s'}(k_1,k_2)$ denotes its Fourier transform,
%The cubic term in the effective potential $V^e$ has the form $\int dk_1 dk_2
%\hat A_{\m, k_1-k_2}\phi^+_{k_1,s,l}\phi_{k_2,s',l}
%\hat \G_{\m,l,s,s' }(k_1,k_2)$; by expanding in Taylor series 
%$\hat \G_{\m,l,s,s' }(k_1,k_2)$ we see that
the {\it anomalous magnetic moment}, corresponding to a term ${Q\over 2m}
\e_{\m,\n} p_\n A_\m \s_{\m\n}$ in the Dirac action,
is obtained from
$
G_{\m,\n}= m \partial_\n \hat \G_{\m,l,R,L}(k_1,k_2)|_0$
while the dressed charge is related to
$\hat\G_{\m,l,s,s}(0,0)$.
%select
%one can derive to {\it anomalous magnetic moment}, which is due to a term  ${Q\over 2m}
%\e_{\m,\n} p_\n A_\m \s_{\m\n}$ in the Dirac action, $p=k_1-k_2$. 

The dressed charge can be expressed by a series
expansion in $\l^2$ with $n$-th coefficients $O(C^n (\l \L/M)^{2 n})$, 
see \cite{M11},  \cite{M12}; the series is therefore convergent 
provided that $\l \L/M$ is small.
There is non-trivial charge renormalization, due to the fact that the Ward Identities are violated at
finite cut-off, and $\ZZ_s$ has to be chosen so that the value of the  dressed charge is just $Q$.
A similar convergent expansion holds for the wave function renormalization, the chiral anomaly
or the 2-point correlations.
Such quantities are associated to terms which are {\it relevant or margina}l in the RG sense, that is 
connected to terms with positive or vanishing scaling dimension ($D=4-{3\over 2}n_\psi-n_A-p$, if $p$ is the order of derivatives in coordinates space). 
They are therefore or directly running coupling constants, as in the case of the dressed charge, or 
with a dominant part depending only on relevant or marginal terms. 

In contrast, the magnetic moment is associated to an {\it irrelevant} term with dimension $D=-1$. The derivative of $\hat \G_{\m,l,s,s'}$  
produces
an extra factor $1/m$, so a naive dimensional estimate for the $n$-th order of
$\partial_\n \hat \G_{\m,l,R,L}$ is $O(m^{-1}  C^n (\l \L/M)^{2 n})$; this is of no use
for estimating the error done truncating the series, as 
explicit computations of lowest orders are
${m\over M^2}$. One needs therefore to improve the dimensional bounds by implementing suitable cancellations at any order in the convergent expansion.
\vskip.3cm
{\bf Theorem.} {\it Given \pref{ippo} with $\L>M$ and $M/m=e^\b$, $\b>0$ 
we can write
$G_{\m,\n}=\sum_{n=1}^\io G_{\m,\n}^{(n)} \l^{2 n}$ with, in the limit $L\to\io$
\be
|G_{\m,\n}^{(n)} |\le  {m^2\over M^2} \b^{2 n} C^{2 n} ({\L^2\over M^2})^{n-1}  (\log {\L\over M} )^{2 n}\label{bo}
\ee 
and $C$ is a constant independent on $M,m,\L$.
}
\vskip.3cm
The above result proves analyticity
of the magnetic moment 
for $\l C \b   (\L/ M) ^{1+0^+}<1$. 
Note the presence of the small factor
${m^2\over M^2}$ in the r.h.s. of \pref{bo}, which is obtained implementing cancellations in the expansion. $C$ is an
$O(1)$ constant whose value can be obtained collecting all constants in the bounds below. In particular it is found 
$G_{\m,\n}={m^2\over M^2} \l^2 (A^\L_{\m\n} + O(\l^{2 } ({\L^2\over M^2})^{1+0^+}))$ with $A^\L_{\m\n}=O(1)$, from which
an upper and lower bound follows (there is no extra $\log {\L\over m}$ in the lowest order).
One can exclude therefore 
non perturbative effects and justify truncation providing
a rigorous estimate of the error. Note also that 
$A^\L_{\m\n}=A^\io_{\m\n}(1+O({M^2\over \L^2}))$
so that  the result is non sensitive to the cut-off for  ${M\over\L}<<1$. Similar considerations can be done for higher order truncation.

The rest of the paper is organized in the following way.
In \S II we
perform an RG integration, in which the main novelty is
that certain irrelevant terms are renormalized to improve
the scaling dimension of the theory thanks to cancellations. In \S III we introduce the tree expansion and we get a
bound for the effective potential. In \S IV we show that the expansion
for the anomalous magnetic factor has suitable cancellations 
allowing to get the bound \pref{bo}. Finally in \S V the
conclusions are presented.

\section{Renormalization Group analysis}

It is convenient to introduce the 
{\it Generating Function} 
\be
e^{W_{\L}(A,\phi)}=\int P(d\psi) e^{V(\psi)+B(A,\psi)}\label{eff} 
\ee
with $B(A,\psi)=\int dx A_\m j^{e.m.}_{\m,x}
+
\int dx 
(\bar\psi_x\phi_x+\psi_x\bar\phi_x)$. 
The two point Schwinger function is
$
S^{\L}_{i,s,s'}(x,y)={\partial^2 W_{\L} \over\partial \phi^-_{x,i,s}\phi^+_{y,i,s'}}|_0$ and the 3-point
is \be S^{\L}_{\m,i,s,s'}(z;x,y)={\partial^3 W_{\L}\over \partial A_{\m,z} \partial \phi^-_{x,i,s}\phi^+_{y,i,s'}}\ee 
The Fourier transform are defined as $\hat S^{\L}_{i,s,s'}(k)$,
$\hat S^{\L}_{\m,i,s,s'}(k_1,k_2 )$
%=\int_{(0,L]^4} dx e^{i k x}  e^{i k y }\G^{L,\L}_{\m,i,s}(0;x,y)$
with $k={2\pi\over L} n$, with $n$ integer vector. 
%We denote 
%$\lim_{L\to\io} \hat \G^{L,\L}_{\m,i,s}=\hat \G^{\L}_{\m,i,s}$ and similar %notation is used for the other quantities.
Using that 
$-V^{e}_\L(A, g*\phi)+(\phi,g*\phi)=W_\L(A,\phi)$
obtained from the change of variables $\psi+g*\phi\to \tilde \psi$ if  $g*\phi=\int dx g(x,y)\phi_y$,
we can write
\be
\hat\G_{\m,l,s,s'}(k_1,k_2)=
\hat g^{-1}_{l,s}(k_1)  \hat S_{\m,l,s,s' }\hat g^{-1}_{l,s'} (k_2)\label{appo}
\ee

We compute the correlations by an exact Renormalization Group analysis.
The cut-off function is written as
\be
\chi_N(k)=\sum_{h=-\io}^N f_h(k)\quad f_h(k)=\chi(\g^{-h} k)-\chi(\g^{-h+1} k)
\ee
so that $f_h(k)$ is a smooth cut-off function selecting momenta
$\g^{h-1}\le |k|\le \g^{h+1}$; we also call 
$\chi_h(k)=\sum_{j=-\io}^h f_j(k)$ the cut-off function selecting momenta $|k|\le \g^h$. 
The generic integration step can be inductively defined in the following way.
If $\VV^{(N)}=V+B$ and assume that we have integrated
the fields $\psi^{(N)},\psi^{(N-1)},...,  \psi^{(h+1)}$, then 
\be
\int P(d\psi) e^{\VV^{(N)}(\psi, A,\phi)}=\int P(d\psi^{(\le h)})e^{\VV^{(h)}(\sqrt{Z_h}  \psi^{(\le h)},A,\phi)}\label{15}
\ee
with $
\VV^{(h)}(A, \psi^{(\le h)})=$
\be
\sum_{l,m=0}^\io \int d\underline x d\underline y \sum_{\underline s, \underline \e}  \; W^{(h)}_{l,m} (\underline x,\underline y))\prod_{j=1}^l 
\psi^{\e_j, (\le h)}_ {s_j, i_j, x_j} \prod_{j=1}^m A_{\m_j, y_j}^{\e_j}\label{nm}
\ee
with $\psi$ including also the $\phi$ fields 
and $P(d\psi^{(\le h)})$ has propagator $g^{(\le h)}_i(x,y)=$
\be
{1\over L^4}\sum_k e^{i k(x-y)}
\chi_h(k) \begin{pmatrix} Z_{h,i}^L  \s^L_\m k_\m & m_{h,i}\\ m_{h,i} & Z_{h,i}^R \s^R_\m k_\m \end{pmatrix}^{-1} \label{prop}
\ee
%
%with $m_{h,i}=\bar m_{h,i}\sqrt{Z_{h,i}^R Z_{h,i}^L }$.
The single scale propagator is bounded by
\be
|g^{(h)}(x)|\le C \g^{3 h} e^{-(\g^h |x|)^{1\over 2}}
\ee
hence $\int dx |g^{(h)}(x)|\le C \g^{-h}$; moreover $\int |v_W(x)|\le C/M_W^2$
and $\int |v_Z(x)|\le C/M_Z^2$.

$\VV^{(h)}$ is sum of monomials of any degree in the fields, with
scaling dimension $D=4-{3\over 2}l-m$. We introduce a renormalization procedure
extracting from $V^h$ not only the terms with scaling dimension $\ge 0$ (that is only the relevant or marginal term), but also the irrelevant terms with scaling dimension $-1$. 

We write therefore
\be
\int P(d\psi^{(\le h)})e^{\LL \VV^{(h)}(\sqrt{Z_h}  \psi^{(\le h)},A,\phi)+\RR \VV^{(h)}(\sqrt{Z_h}  \psi^{(\le h)},A,\phi) }
\ee
where $\RR=1-\LL$ is the renormalization operation and $\LL$ acts on the monomials in $V^h$ in the following way
\bea
&&\LL \hat W_{2,1;s,s}(k_1,k_2)=\hat W_{2,1;s,s}(0,0)+\nn\\
&&k_1 \partial_1 \hat W_{2,1;s,s}(0,0)+k_2 \partial_2 \hat W_{2,1;s,s}(0,0)\nn\\
&&\LL \hat W_{2,1;L,R}(k_1,k_2)=\hat W_{2,1;L,R}(0,0)+\\
&&k_1 \partial_1 \hat W_{2,1;L,R}(0,0)|_{m=0}+k_2 \partial_2 \hat W_{2,1;L,R}(0,0)|_{m=0} \nn\\
&&\LL \hat W_{2;s,s}(k)=\hat W_{2;s,s}(0)+k 
\partial \hat W_{2;s,s}(0)+{1\over 2} k^2 \partial^2 \hat W_{2;s,s}(0)\nn\\
&&\LL \hat W_{2;L,R}(k)=\hat W_{2;L,R}(0)+\nn\\
&&k \partial \hat W_{2;L,R}(0)+{1\over 2} k^2 \partial^2 \hat W_{2;L,R,s}(0)|_{m=0}\nn
\eea
Note that the propagators involving the same chirality are odd in the exchange $k\to-k$  
and the ones involving different chitalities are even. Therefore 
\be
\hat W_{2;s,s}(0)=\partial^2 \hat  W_{2;s,s}(0)=0\ee 
as
is given by graphs with an odd number of diagonal propagator and an even of non diagonal ones; 
$\partial \hat W_{2;L,R}(0,0)=0$ as is given by graphs with an even number of diagonal propagators and an odd of non diagonal ones; 
$\partial^2 \hat W_{2;L,R}(0,0)_{m=0}=0$ as they require a non diagonal propagator to be non-vanishing;
$\hat W_{2,1;L,R}(0,0)=0$ as there is an odd number of non diagonal propagator and an odd of diagonal ones; 
$\partial \hat W_{2,1;L,R}(0,0)_{m=0}=0$ as there 
as they require a non diagonal propagator to be non-vanishing;$\partial \hat W_{2,1;s,s}(0,0)=0$ as is given by graphs with an odd number of diagonal propagators and an even of non diagonal ones.

We can write therefore
\be
\int \tilde P(d\psi^{(\le h)})e^{\tilde\LL \VV^{(h)}(\sqrt{Z_{h-1}}  \psi^{(\le h)},A,\phi)+\RR \VV^{(h)}(\sqrt{Z_{h-1}}  \psi^{(\le h)},A,\phi) }\label{al}
\ee
where for $h\le N-1$
\be
\LL V^h(\sqrt{Z_{h-1}}  \psi^{(\le h)},A,\phi)=\sum_{s} \int dx 
Z^A_{h,s}
A_\m  \psi^+_{x,l,s} \s_\m^s \psi^+_{x,l,s}
\ee
and $\tilde P(d\psi^{(\le h)})$ has propagator given by \pref{prop} with $Z_{h,i,s}$ replaced by $Z_{h-1,i,s}=Z_{h,i,s}+\partial W^h_{2,s,s}(0)$,
$m_{h-1}=m_h+W^h_{2,R,L}(0)$ and $Z^A_{h-1,s}={Z_{h-1,l,s}\over Z_{h,l,s}}(Z^A_{h,s}+W_{2;1,s,s}(0,0))$. In the case of $\psi\phi$ or $A\psi \phi$ we use the fact that the $\LL$ part is vanishing as there is surely a propagator $g^h(0)=0$, hence there is no running coupling constant associated.

Using that $\tilde P(d\psi^{(\le h)})=P(d\psi^{(\le h-1)})P(d\psi^{(h)})$ with
$g^{(h)}$ given by
\pref{prop} with $\chi_h$ replaced $f_h$, we get that \pref{al} can be written as the r.h.s. of \pref{15}
with $h-1$ replacing $h$ and
\be
\VV^{h-1}=\sum_{n=0}^\io {1\over n!}\EE^T_h(\tilde\LL \VV^{(h)}+\RR \VV^{(h)};...; \tilde\LL \VV^{(h)}+\RR \VV^{(h)})\label{ipp} 
\ee
where $\EE^T_h$ are the fermionic truncated expectations, that is $\EE^T_h(O;n)={\partial^n\over \partial \l^n}\log \int P(d\psi^h)e^{\l O}|_0$.
\insertplot{290}{60}
{\ins{60pt}{27pt}{$+$}
\ins{140pt}{27pt}{$+$}
\ins{220pt}{27pt}{$...$}
}
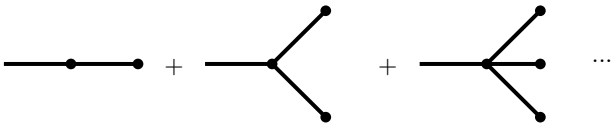
{figjsp44abbc}{\label{n11} Graphical representation of \pref{ipp}, that is  $\EE^T_h(\tilde \VV^{(h)})+
 {1\over 2}\EE^T_h(\tilde \VV^{(h)};\tilde \VV^{(h)})+ {1\over 3!}\EE^T_h(\tilde \VV^{(h)};\tilde \VV^{(h)};\tilde \VV^{(h)})+...$ with 
$\tilde\VV^{(h)}=\tilde\LL \VV^{(h)}+\RR \VV^{(h)}$
}{0}
The procedure can be then iterated up the scale of the fermionic mass defined as 
\be
\g^{h^*}=m_{h^*}\label{ref}
\ee
At this point one can write $\int P(d\psi_l^{\le h^*}) e^{V^{h^*}
(\psi_l,\psi_\n,A,\phi)}=e^{V^{h^*}(\psi_\n,A,\phi)}$ using that $
|g^{\le h^*}(x)|\le C \g^{3 h^*} e^{-(\g^{h^*} |x|)^{1\over 2}}$.
The  integration of the remaining scales is done as above, the only difference that only the fields $\psi_\n$ remain; the $\psi_l$ has been already 
integrated out. Note that $m_{h,\n}=0$ by symmetry.

In order to write explicitly the effective potential $\VV^{(h-1)}$
one has to express the $\RR \VV^{h}$ in the r.h.s. of \pref{ipp}
in terms of sum of truncated expectations, while no further expansion is done in the $\LL\VV^h$; a graphical representation of a term is in fig. 2.
\insertplot{990}{60}
{
\ins{140pt}{27pt}{$$}
\ins{220pt}{27pt}{$$}
}
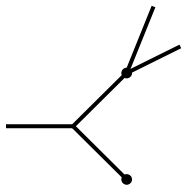
{figjsp44abbd2}{\label{n11} Graphical representation of
${1\over 2}\EE^T_h
(\tilde\LL \VV^{(h)}; \RR{1\over 2} \EE^T_{h+1} (\VV^{h+1}; \VV^{h+1}))$
}{0}

This procedure can be iterated up to the scale $N$, resulting in a tree expansion described below.

\section{Renormalized Expansion}

Iterating \pref{ipp} we get an expansion for $\VV^{h}$ in terms of 
{\it trees} \cite{G}, see Fig. 3 
\be
\VV^{(h)}(\psi^{(\le h)},A,\phi) =
\sum_{n=1}^\io\sum_{\t\in\TT_{h,n}}
\VV^{(h)}(\t)\;,\ee
with $\t$ a tree, constructed
by joining a point, the root $r$, with an ordered set of $n\ge 1$
end-points, and associating a label $h\le N-1$ with the root; 
moreover  
we introduce a family of vertical lines, labeled by an integer taking values
in $[h,N+1]$ intersecting all the non-trivial vertices,  the endpoints and other points called trivial vertices. 
\insertplot{350}{140}
{\ins{120pt}{100pt}{$v$}
\ins{40pt}{90pt}{$r$}
\ins{60pt}{90pt}{$v_0$}
\ins{60pt}{-5pt}{$h+1$}
\ins{100pt}{90pt}{$v'$}
\ins{120pt}{-5pt}{$h_v$}
\ins{235pt}{-5pt}{$N$}
\ins{255pt}{-5pt}{$N+1$}
}
{treelut2}{\label{n11} A labeled tree 
}{0}
To each vertex $v$ is associated a scale $h_v$;
they are partially ordered
and, if $v_1$ and $v_2$ are two vertices and $v_1<v_2$, then
$h_{v_1}<h_{v_2}$; moreover given $v$ there are $S_v$ points following $v$. The first vertex has scale $h+1$.
The end-points $v$ can be 1)$\l$ points to which is associated $V(\psi)$, or $\phi$ points to which is associated $B(\psi,0,\phi)$
; in this case the scale is $h_v=N+1$. 2) $Z$ points, and are associated to
$\LL\VV^{h_v-1}(\psi^{(\le h_v-1)},A)$ and in this case the scale is $h_v \le N+1 $ 
and there is the constraint 
that $h_v=h_{v'}+1$ if $v'$ is the non trivial vertex preceding $v$.  
Given a vertex $v$
we call $m^\l_v$ the number of $\l$ points following $v$, $m^\phi_v$
the number of $\phi$ points following $v$ and $m^A_v$ the number of $Z$ points following $v$.

With the above definitions the value of $\VV^{(h)}(\t)$ is obtained iteratively by the relations
\be
\VV^{(h)}(\t)={(-1)^{s+1}\over s!} \EE^T_{h+1}[\bar
\VV^{(h+1)}(\t_1);..; \bar
\VV^{(h+1)}(\t_{s})]\label{ell}
\ee
where $\t_1,..,\t_s$ are the subtrees with root in $v$, 
$\bar
\VV^{(h+1)}(\t)=\RR \VV^{(h+1)}(\t)$ if
the subtree $\t_i$ contains more then one end-point,
while if $\t_i$ contains only one end-point $\bar
\VV^{(h+1)}(\t)$ is
$V(\psi^{(\le N)},0,\phi)$ if 
$h=N$ or if  $h\le N$
is
$\LL\VV^{h+1}(A, \psi^{(\le h+1)})$.

By \pref{ell} we see that 
$\VV^{(h)}(\t)=\sum_P \VV^{(h)}(\t,P)$, where $P$ is the set of all $P_v$ associated to the vertices of the tree, corresponding to subsets of the labels of the fields associated to the end-points following $v$. We call $V$ the vertices such that $P_v$ is different with respect to the preceding one.

The $\VV^{(h)}(\t,P)$ can be represented as sum of
renormalized Feynman graphs. The difference with respect to the usual Feynman graphs is that the scale labels of the tree $\t$, corresponding to vertices 
$v\in V$, can be represented as a set of clusters enclosing the end-points. To each point is associated an element of $V$ or $\LL V^h$,
represented graphically as a point with half lines to be contracted.
To each line is associated a scale, and there is the constraint that all the lines inside a cluster $v$ have scale $\le h_v$, and at least one of them is at scale $h_v$.
The $\RR$ operation is applied on the clusters depending on the number of the external lines.
\insertplot{900}{190}
{}
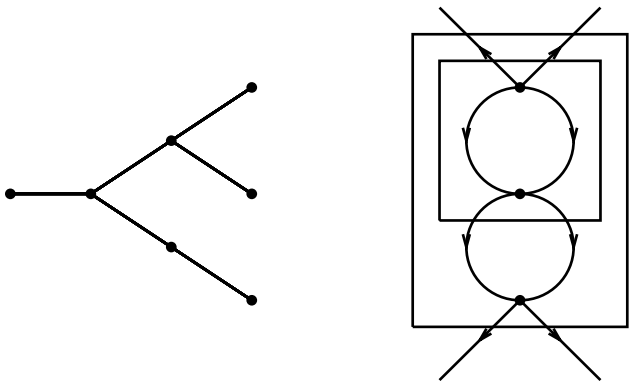
{fig9a}{\label{n11} A graph with its clusters and the corresponding tree; the smaller cluster has scale $h_1$ and the larger $h+1$.
}{0}

Each graph is finite but one needs that 
the sum over the scale labels is finite. 
Let us consider for instance the graph in Fig. 4;
One can bound the sum over the scales by, up to a constant, if $N\ge h_1 \ge h+1$, $({\l^2\over M^2})^3
\sum_{h_1}\g^{2h_1}\g^{2h }$. In this example, there is no $\RR$ operation in the subgraphs. In contrast, the $\RR$ operation is present in the graph in  Fig 5. 

\insertplot{590}{100}
{}
{figjsp44aaaa2}
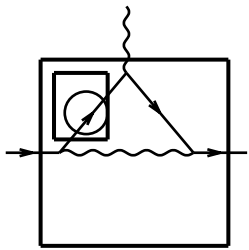
{\label{h2} A graph requiring renormalization; the smallest cluster has scale $h_1$ and the larger $h+1$.
} {0}
%
%\insertplot{300}{200}
%{}
%{fig9a}{\label{n11} A graph with its clusters and the corresponding tree
%}{0}
The effect of the $\RR$ operation can be written as
\be
\RR \hat W^{(h_{v})}_{2;s,s}(k)=k^3 \int_0^1 dt \partial^3 \hat W^{(h_{v})}_{2;s,s}(t k)\ee
Therefore the effect of $\RR$ is to produce an extra $k^3$; to the external lines of $\hat W^{(h_{v})}$
is associated  
a propagator $g^{(h_{v'})}(k)$, if $v'$ is the vertex $\in V$ following $v$,
with a cut-off function restricting the 
value of $k$ to $\sim \g^{h_{v'}}$. Similarly the derivatives on  $\hat W^{(h_{v})}$
are applied on propagators with scale $\ge h_v$. Therefore the effect of the $\RR$ operation is to produce an extra
$\sim \g^{3(h_{v'}-h_v)}$ factor. 
Regarding the terms 
$W_{2;R,L}(k)$, in addition to such term there is also a contribution of the form
\be 
{1\over 2} k^2 \partial^2 W^{h_v}_{2;L,R}(k)|_{m=0}-{1\over 2} k^2 \partial^2 W^{h_v}_{2;L,R}(k)
\ee
Such terms are present only for $i=l$; by phase symmetry when $i=\n$ there are no terms 
$L,R$, by invariance under $\psi_{s,\n}\to e^{i\a_s}  \psi_{s,\n}$.
The bound for the propagator involving $L$ and $R$ fields has 
an extra factor ${m_{h_v}\over \g^{h_v}}$
and, for $h_v\ge h^*$ \pref{ref}
\be
{m_{h_v}\over \g^{h_v}}={m_{h_v}\over m_{h^*}}{m_{h^*}\over \g^{h_v}}
\le  \g^{h_{v'} -h_v}
\ee
as the smallest scale of the external fields of type $i=l$ is $h^*$, that is $h_{v'}\ge h^*$.

Without the $\RR$ operation, the graph in Fig 5 is bounded by, $h_1\ge h_2$
$({\l^2\over M^2})^3
\sum_{h_1}\g^{5h_1}\g^{h}$; the $\RR$ operation adds an extra 
$\g^{3(h- h_1)}$.

A finite bound for any graph is still not sufficient for getting convergence; the number of graphs has a factorial growth.
Therefore it is convenient 
to represent
the truncated expectation as \cite{B} 
$
\EE^T_{h}(\tilde\psi^{(h)}(P_1);\tilde\psi^{(h)}(P_2);
...;\tilde\psi^{(h)}(P_s))=$
\be \sum_{T}\prod_{l\in T} g^{(h)}(x_l-y_l)
\int dP_{T} \det
G^{h,T}\label{lap}\ee
where $\tilde\psi^{(h)}(P)$ are monomials in the $\psi$,
$
T$ is a set of lines forming an {\it anchored tree graph}
connecting the set $P_1,..,P_s$ and $\det
G^{h,T}$ is a matrix containing the fields not belonging to $T$.
The crucial point is that, using Gram inequality, the 
$\det
G^{h,T}$ can be bounded by a constant times the number of fields.
This is a way to implement the well known fact that fermionic series expansion can be convergent, in contrast to bosonic ones.
Using \pref{lap} one gets for a class of graph with a chosen tree $\t$ and $P$
the same bound as for a single graph, without factorials.
If $||V(\t,P)||$ denotes the integral of the modulus over all the coordinates except one, then, see eg \cite{M12},
$
||V(\t,P)||\le$
\be
 C^n \prod_{v \in V } \g^{4 h_v (S_v-1)} \g^{-3 h_v n_v} \prod_{v \in V } \g^{z_v(h_{v'}-h_v  )}   (\l^2/M^2)^n
\ee
where 
$n_v$ is the number of propagators in the cluster $v$ and not in any smaller one, $v'$ is the vertex in $V$ preceding $v$ and $S_v$ are the vertices following $v$ (or the maximal clusters in $v$).
The factor $\prod_v \g^{z_v(h_{v'}-h_v  )}$ is the effect of the renormalization procedure;
$z_v=3$ in the term $\psi\psi$ or $\psi\phi$ and $z_v=2$ in the terms 
$A\psi\psi$ or $A\psi\phi$. 
We use the relations
$
\sum_{v\in V} (h_v-h)(S_v-1)=\sum_{v\in V}  (h_v-h_{v'})(m^\l_v+m^A_v+m^\phi_v-1)
$
and $\sum_{v\in V} (h_v-h) n_v=\sum_{v\in V}  (h_v-h_{v'})(2 m^\l_v+m^A_v+m^\phi_v-n_v^e/2)$
where $n^e_v$ is the number of external $\psi,\phi$ lines from the cluster $v$. Therefore the bound becomes , if $\bar D_v=
4-3 n_v^e/2+ 2m^\l_v-m^A_v-m^\phi_v$,
$
||V(\t,P)||\le$
\be
 C^n  \g^{(4-3/2 l+2 m^\l-m^A-m^\phi )h }
\prod_{v \in V} \g^{(h_v-h_{v'})(\bar D_z-z_v)}(\l^2/M^2)^n
\ee
where $l$ are the external $\psi,\phi$ lines associated to $V(\t,P)$ and $z_v=3$ for the $v$ with 2 external $\psi$ lines
and $z_v=2$ for the  $v$ with 2 external $\psi$ lines
and one $A$ line. 
We use now the relation, $i=\phi, \l$
$
\g^{h m^i_{v_0}}   \prod_{v\in V}\g^{(h_v-h_{v'})m^i_v}
=\prod_{v\in V}\g^{h_{v}\bar m^i_v}
$, 
where  $\bar m^i_v$ is the number of endpoints of type $i$ contained in
$v$ and not in any smaller cluster.
% (or the end-points $v$ such that $v'=v$).
Therefore if $D_v=
4-3 n_v^e/2-m_v^A$
\bea
&&||V(\t,P)||\le 
C^n  \g^{h (4-3/2 l-m^A )}
\prod_{v \in V} \g^{(h_v-h_{v'})(D_{v}-z_v)}\nn\\
 &&[\prod_{v\in W_\l} \g^{2 (h_{v^*}-N) } (\l^2 \g^{2 N}/M^2)^n  [\prod_{v\in W_\phi}  \g^{- h_{v^*}}]\label{sap}
\eea
where $W_\l$ and $W_\phi$ are the endpoints of $\l$  or $\phi$ type
and $v^*$ is the first non trivial vertex preceding $v$.
Note that if $v\in W_\phi$ then 
$ h_{v^*}=h_k, h_k+1$; the reason is that the corresponding contribution is of the form $g^{h_{v^*}}(k) W$ hence is non-vanishing only for such scales.

We consider first the contribution to the effective potential when there are no $\phi$ end-points.
The scale $h$ is
fixed so that the sum
over all the possible scales can be done summing over all the possible scale differences (the scale $h$ is fixed) hence, 
if $\tilde D_v=D_v-z_v\ge 2$, 
\be
\sum_{\{h\}}\prod_v
\g^{(h_v-h_{v'})\tilde D_v}\le C^n \prod_v\g^{-|n_v^e|/4}(\sum_{q=1}^\io \g^{-2 q})^{4 n}\ee
as $-\tilde D_v-\chi(n^e_v\ge 8) |n_v^e|/4\ge 2$. The factor $\g^{-|n_v^\psi|/4}$
is used to sum over $P$.
Then $\sum_\t \sum_P  |V(\t,P)|\le C^n  \g^{h(4-3/2 l-m ) } (\l^2 \g^{2 N}/M^2)^n$
implying summability over $n$ if $(\l^2\g^{2 N}/M^2)$ is small enough.

As an example, the bound \pref{sap} for the graph in Fig.4 is
given by, up to the factor $({\l^2\g^{2 N} \over M^2})^3$
$\g^{- 2 h}\sum_{h_1}\g^{-2(h_1-h)} \g^{4(
h_1-N)} \g^{2(h-N)}$. Similarly the bound for the graph in Fig 5
is
$\sum_{h_1}\g^{-(h-h_1)} \g^{3(h- h_{1}))} \g^{4(h_1-N)}\g^{2(h-N)}$.

\section{The anomalous magnetic moment}

The 3-point function $S^{\L}_{\m,l,s,s'}(z;x,y)$
with external fields of type $l$, can be written as
$
S^\L_{\m,l,s,s'}= \sum_\t \sum_P S_\m(\t,P)$
where the sum is over all the trees with two $\phi$ end points and a $Z$ end point.
We choose the momentum of the
external fermionic lines 
as $|k_1|,|k_2|\le \g^{h^*}$. 
\insertplot{200}{100}
{\ins{26pt}{40pt}{+}\ins{130pt}{40pt}{+}\ins{240pt}{40pt}{+...}
}
{figjsp44aaaa1}
{
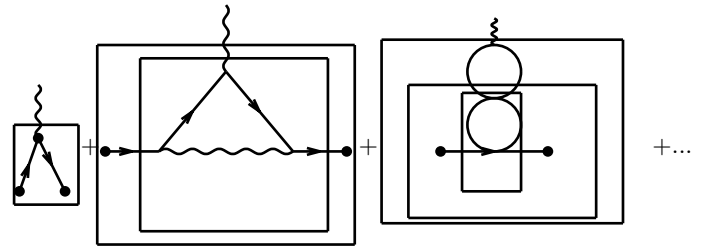
\label{h2} Some graphs contributing to the  functions $\G$; the $\phi$ lines are not represented (they are meant as external to all clusters). In the first the scale of the cluster is $h^*$; in the second the smallest has scale $h_1$ and the larger $h^*$; in the third the smaller $h_1$, the medium $h^*$ and the larger $h$.
} {0}
We can distinguish between trees with no $\l$ end-points and at least a $\l$ end-point. In the first case one has only a contribution to 
$\hat S^\L_{\m,l,s,s'}(k_1,k_2)$
of the form, see the first graph in Fig 6
\be
\sum_{\bar s} Z^A_{h^*,\bar s} g_{\bar s,s}^{(\le h^*)}(k_1)\s^{\bar s}_\m g_{\bar s,s'}^{(\le h^*)}(k_2)
\label{appo1}
\ee 
In the second case there is at least a $\l$ end-point, like in the second and third graph in Fig. 6. Let us consider the smallest cluster $v_s$ containing the point $v_\phi$ associated to the $\phi$
end-point; note that $h_{v_s}=h^*$ as the momentum of the external lines is assumed $\le \g^{h^*}$, and 
the contraction of the $\psi$ field produces
a propagator with the same momentum as the external one.
In the cluster $v_s$ there is also surely a $\l$ end-point point $\tilde v_\l$ (there is at least either a $\l$ or $Z$ point, and there is only one $Z$),
and $\tilde v_s$ is the smallest cluster containing $\tilde v_\l$ contained in $v_s$. 

\insertplot{250}{100}
{\ins{30pt}{50pt}{$v_0$}\ins{42pt}{57pt}{$v_1$}
\ins{49pt}{87pt}{$v_{s+1}$}
\ins{40pt}{75pt}{$v_s$}\ins{70pt}{65pt}{$v_\phi$}
\ins{60pt}{95pt}{$\tilde v_s$}\ins{85pt}{105pt}{$\tilde v_\l$}
}
{figjsp467}
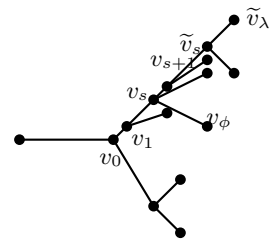
{\label{h2} A tree with the path from $\tilde v_\l$ to $v_0$
} {0}

In terms of trees, there is a path from $\tilde v_\l$ to $v_0$, $v_0$ being the first vertex belonging to $V$ (or a sequece of clusters
one enclosed in the other) 
$v_0 < v_1<..v_s<v_{s+1}.. <\tilde v_s$, an $h_{v_0}\le...h_{v_s}<...h_{\tilde v_s}$ with $h_{v_s}=h^*$, see Fig 7.

We get therefore the bound, if $\tilde D_v=D_v-z_v\le -2$
$||S_\m(\t,P)||\le$
\be
 C^n \prod_{v\in V} \g^{(h_v-h_{v'})\tilde D_v }
\g^{2 (h_{\tilde v_s}-N) } (\l^2\g^{2 N}/M^2)^n  \g^{-2 h^*}\label{37}
\ee
which implies 
\bea
&&||S_\m(\t,P)||\le
 C^n \prod_{v\in V} \g^{(h_v-h_{v'})(\tilde D_v+\th_v) }\g^{\th (h_{v_0}-h_{v_s} )}\nn\\ 
&&\g^{\th (h_{v_s}-h_{\tilde v_s} )} \g^{2 (h_{\tilde v_s}-N) } (\l^2\g^{2 N}/M^2)^n  \g^{-2 h^*} 
\eea
where 
$\th_v=\th<2$ for $v=v_0 ,v_1,..,v_s..,\tilde v_s$ and $\th_v=0$ otherwise. 
\insertplot{300}{100}
{\ins{30pt}{50pt}{$v_0$}\ins{40pt}{75pt}{$v_s$}\ins{70pt}{65pt}{$v_\phi$}\ins{70pt}{80pt}{$v_\phi$}
\ins{60pt}{95pt}{$\tilde v_s$}\ins{85pt}{105pt}{$\tilde v_\l$}
\ins{85pt}{85pt}{$\tilde v_\l$}\ins{40pt}{30pt}{$v_A$}
}
{figjsp467a}
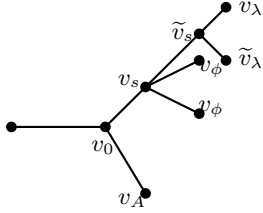
{\label{h2} The tree corresponding to the third graph in fig 6; $h_{v_0}=h$, $h_{v_s}=h^*$,  $h_{\tilde v_s}=h_1$.  
} {0}
Therefore
\bea
&&||S_\m(\t,P)||\le 
\g^{\th (h_{v_0}-h^*)}\g^{\th (h^*-N) }\\
&&
 C^n \prod_{v\in V} \g^{(h_v-h_{v'})(\tilde D_v+\th_v) }\g^{\th (h_{v_0}-h_{v_s})}\g^{-2 h^*} (\l^2\g^{2 N}/M^2)^n\nn
\eea
The sum over all the scale difference is done again using the factors $\g^{(h_v-h_{v'})(\tilde D_v+\th_v) }$;
the sum over $h_{v_0}$ is controlled by the factor $\g^{\th (h_{v_0}-h^*)}$ hence the bound 
for the contributions to $\hat S^\L_{\m,l,s,s'(k_1,k_2)}$
with at least a $\l$ term 
is 
\be
\sum_\t \sum_P ||S(\t,P)||\le C^n  \g^{-2 h^*} \g^{\th(h^*-N)}
 (\l^2\g^{2 N}/M^2)^n
\ee
Note that such terms
are subdominant due to the an extra factor $\g^{\th(h^*-N)}$.
The 3-point function is equal to the free one with renormalized parameters, up to more regular terms containing at least an irrelevant $\l$ interaction.
A similar result holds for the 2-point function.

As an example,  the second graph in Fig 6 is bounded by (up to a factor
$(\l^2\g^{2 N} /M^2)$)
$\g^{-2 h^*}  \sum_{h_1\ge h^*}\g^{2 (h^*-h_1)}\g^{2 (h_1-N)}$  where the factor 
$\g^{2 (h^*-h_1)}$ is produced by the $\RR$ operation: 
hence it is bounded by
$ \g^{-2 h^*}\g^{\th(h^*-N)}$.

The third graph gives (up to a factor
$(\l^2\g^{2 N} /M^2)^2$), for $h_1\ge h^*\ge h$
$\g^{-2 h^*} \sum_{h_1,  h}\g^{2h}\g^{2h_1} \g^{-4N}$ which can be written as
$\g^{-2 h^*}\sum_{h_1, h}\g^{2(h-h^*)}\g^{2(h^*-h_1)} \g^{4 (h_1-N)}$ and finally by 
$\g^{-2 h^*}\g^{\th (h^*-N)}$.
Finally for the graph in Fig 5 (if the external lines are propagators, and there is an extra $\RR$)
, one gets for $h_1 \ge h\ge h^*$
 $\g^{-2 h^*} 
\sum_{h_1 ,h} 
\g^{2 (h^*-h)}\g^{2(h- h_{1}))} \g^{4(h_1-N)}\g^{2(h-N)}$
which is surely smaller than $\g^{-2 h^*} \g^{\th(h^*-N)}$.

We arrive finally to the bound for $G_{\m\n}= m \sum_{\t,P}  \partial \G_A(\t,P)$. By \pref{appo},
\pref{appo1} we see that the contribution from the dominant term to $S_\m$
as the derivative is vanishing. One has therefore to consider the derivative of the amputated contributions with at least a $\l$ end-point. Moreover we consider the term with $s\not=s'$, which are the only contributing at zero momentum.
We get 
$
m ||\partial \G_A(\t,P)||\le$
\be m^2 \g^{-h_{v_0}} \g^{-h^*}
 C^n \prod_{v\in V} \g^{(h_v-h_{v'})\tilde D_v }
\g^{2 (h_{\tilde v_s}-N) } (\l^2\g^{2 N}/M^2)^n  
\ee
where with respect to \pref{37} there is an extra
$\g^{-h_{v_0}}$ from the derivative and a missing $\g^{-2h^*}$.
The above expression 
can be rewritten as
\bea
&&m || \partial \G_A(\t,P)||\le
m^2 \g^{-h_{v_0}} \g^{-h^*} \g^{{3\over 2}(h_{v_0}-h^*)}
\\
&&\g^{2(h^*-h_{\tilde v_s} )}
 C^n \prod_{v\in V} \g^{(h_v-h_{v'})(\tilde D_v+\bar\th_v) }
\g^{2 (h_{\tilde v_s}-N) } (\l^2\g^{2 N}/M^2)^n  \nn
\eea
where $\bar\th_v=3/2$ for $v=v_0 ,v_1,..,v_s$
and $\bar\th_v=2$ for $v=v_{s+1},..,\tilde v_s$,  
and $\th_v=0$ otherwise. 
Therefore
\be
\g^{-h_{v_0} } \g^{-h^*}\g^{3/2(h_{v_0}-h^*) }\g^{2 (h^*-N) } 
\le \g^{1/2(h_{v_0}-h^*) }\g^{-2N }
\ee
and finally we get
\bea
&&m || \partial\G_A(\t,P)|| \le\\
&&
\g^{1/2(h_{v_0}-h^*) }m^2 \g^{-2N }
C^n \prod_{v\in V} \g^{(h_v-h_{v'})(\tilde D_v+\bar\th_v) }
 (\l^2\g^{2 N}/M^2)^n  \nn
\eea
Now the sum over $h_{v_0}$ is done using the factor 
$\g^{1/2(h_{v_0}-h^*) }$; the sum over the difference of scales is done using 
$\g^{(h_v-h_{v'})(\tilde D_v+\bar\th_v) }$ for any $v$
except the ones between $v_s$ and $\tilde v_s$;  there are at most $2n$ of such vertices so that their an extra factor  
$|N-h^*|^{2 n}$. The same over $P$ is done in the same way as we can extract a factor 
$\g^{-n^e_v/6}$ for $n^e_v\ge 8$.
Therefore the bound is, $n\ge 1$ 
\be
m || \partial \G_A(\t,P)|| \le {m^2\over M^2}\l^2 (\log (m/\g^N))^{2 n}
 (\l^2\g^{2 N}/M^2)^{n-1}
\ee
For instance the bound for the second graph in Fig 6 is, obtaining 
$\g^{-2 h^*}$ by the derivatives,
$m^2{\l^2\over M^2} \sum_{h_1\ge h^*} 1$
hence  it is bounded by 
$|h^*-N|{m^2 \l^2\over M^2}$.
In the case of the third graph in Fig 6 for $h_1\ge h^*\ge h$
it can be written as, up to $m^2(\l^2\g^{2N} /M^2)^2$
 $\sum_{ h_1, h}\g^{-h^*}\g^{-h} \g^{2(h-h^*)}\g^{2(h^*-h_1)} \g^{4 (h_1-N)}$;
moreover we can write
$\g^{-h^*+h_1}\g^{-h+h_1}$ as $\g^{-2(h^*-h_1)}\g^{-h+h^*}$ so that we get
 $\sum_{ h_1, h}\g^{(h-h^*)}\g^{ -2N}$
which is bounded by
$|N-h^*|\g^{ -2N}$. 
The contribution of the graph in Fig 5 is 
$h_1 \ge h\ge h^*$
 $\g^{-2 h^*} 
\sum_{h_1 ,h} 
\g^{2 (h^*-h)}\g^{2(h- h_{1}))} \g^{4(h_1-N)}\g^{2(h-N)}$
bounded by $|N-h^*|^2 \g^{-2 N}$ times $m^2(\l^2\g^{2N} /M^2)^3$.

Moreover $(\log \L/m)^{2 n}=(\log \L/M+\log M/m)^{2 n}$ which
is equal to $\sum_p {2 n!\over p! (2n-p)!} (\log \L/M)^{p}
(\log M/m)^{2 n-p}\le  (\log \L/M)^{2 n}
(\log M/m)^{2 n} 2^{2 n}$.

The lowest order contribution to the magnetic moment is given by the second graph in Fig. 6 and the difference between the finite and infinite $\L$
is bounded by (the $\RR$ disappear with the derivative) $m^2\int_\L^\infty dk  {1\over k^4} {1\over k^2+M_Z^2}$ which is $O({m^2\over M_Z^2} {M_Z^2\over \L^2} ) $; this follows from the non-locality of the interaction which was not used in the bounds.

Finally
we have to study the flow of the running coupling constants. 
They verify recursive equations in which there is at least a $\l$ end-point so that proceeding as above we can write 
${Z_{i,s,h-1}\over Z_{i,s,h}}=1+\b^h_z$ with $\b^h_z=O( \g^{\th(h-N)}\  (\l^2\g^{2N}/M^2)^2)$.
This implies that $\lim_{h\to\io} Z_{i,s,h}=Z_{l,s}$ is finite and  $Z_{i,s}=1+O(\l^2\g^{2N}/M^2)$; the fermionic wave function renormalization depends on the particle and chiral index. In the same way  
$\lim_{h\to\io} Z^A_{s,h}=Z_{s}^A$ with
$Z_{s}^A=1+O(\l^2\g^{2 N}/M^2)$ and $m_{h^*}=m(1+O(\l^2\g^{2 N}/M^2))$. Note that $Z^A$ and $Z$ are different,
due to violation of Ward Identities due to momentum regularization
which produces an extra term in the Ward Identities
\be
p_\m \tilde \G^{\L}_{\m,l,s}(k,k+p)=
Q (S^\L_{l,s,s}(k)-S^\L_{l,s,s} (k+p))+\d \G^\L_{l,s}
\ee
where $\tilde \G^{\L}_{\m,l,s}$ is defined as $\G^{\L}_{\m,l,s}$ with $\ZZ_{l,s}=1$, and  
$\d \G^\L_{l,s}$ is similar to $\tilde \G^{\L}_{\m,l,s}$ with the current replaced by
$
\d j_l =\sum_{s} Q \int dk dp C(k,p) \bar\psi_{k,l,s}  \s^s_\m \psi_{k+p,l,s}
$
where $C(k,p)=\not k(\chi^{-1}(k)-1)-(\not k+\not p)(\chi^{-1}(k+p)-1)$.
We have therefore to choose $\ZZ_s$ to impose 
\be
Z_{s}^A/Z_{l,s}=1\ee With this condition the effective potential has the form 
$\int \hat A_{\m, k_1-k_2}Z_{s,i}^{-1/2}  Z_{s',i}^{-1/2}\bar \phi_{k_1,s,i}\phi_{k_2,s',i}
\hat V_{\m,i,s }(k_1,k_2)$ and
$
V_{\m,i,s,s'}=Z^{1/2}_{s,i} Z^{1/2}_{s',i}\hat \G_{\m,i,s,s' }(k_1,k_2) $
with $V_{\m,l,s,s }(0,0)=Q$ and the magnetic moment obtained by
$Z^{1/2}_{s,i} Z^{1/2}_{s',i} G_{\m,i,s,s' }$.

\section{Conclusions}

The series for the magnetic moment is expected to be non convergent and even not asymptotic, and this makes unclear how to evaluate the error introduced by truncation. We consider a  non-perturbative framework  expressing the magnetic moment in terms of Euclidean functional integrals 
with a finite ultraviolet cut-off, considering a Fermi description for Weak forces.
The fact that the
magnetic moment is associated to an irrelevant quantity
in the RG sense requires 
careful estimates
and the implementation of previously unknown cancellations. 
We get that the magnetic moment is expressed by
series which are analytic for $\l  ({\L^2\over M^2})^{(1+0^+)}$ small, 
with relative error due to truncation at order $n$ 
$O(\l^{2(n-1)} ( {\L^2\over M^2})^{(n-1)(1+0^+)})$.
In addition the lowest order coincides with its $\L\to\io$ limit
up to an error terms $O({M^2\over\L^2})$. This excludes non-perturbative phenomena in the regime of parameters where such two errors are small,
that is $\l^2<< {M^2\over\L^2}<<1$, and it justifies the validity of truncation of the series expansion with no cut-off in this regime.

An important open question is if a similar approach
based on rigorous RG
and 
Euclidean functional integrals with cut-off can be repeated keeping the gauge interaction (and loosing analyticity), and if an estimate for the relative
error due to the truncation is obtained with a weaker logarithmic dependence, that
is $O(\l^{2(n-1)} (\log {\L^2\over M^2})^{(n-1)})$, up to a constant depending on the order with some factorial.  
This could allow to include larger and more realistic values of the coupling.
Technical difficulties to be solved are however the need of
a non perturbative control of functional integrals with massless boson and fermions, which
requires an extra decomposition in the gauge boson fields and the fact that one cannot expand in the coupling,
and the understanding of the interplay of the 
anomaly cancellations in a non-perturbative setting.

\end{document}